\documentclass[aps,prfluids]{revtex4-2}

\usepackage{graphicx}
\usepackage{caption}
\usepackage{subcaption}
\usepackage{dcolumn}
\usepackage{gensymb}
\usepackage{textcomp, gensymb}
\usepackage{bm}
\usepackage{amsmath}
\usepackage{physics}

\usepackage[utf8]{inputenc}
\usepackage[T1]{fontenc}

\begin{document}

\title{Effect of local flow geometry on particle pair dispersion angle}

\author{B.L.~Espa\~nol}
\affiliation{Universidad de Buenos Aires, Facultad de Ciencias Exactas y Naturales, Departamento de Física, Ciudad Universitaria, 1428 Buenos Aires, Argentina}
\affiliation{CONICET - Universidad de Buenos Aires, Instituto de F\'{\i}sica Interdisciplinaria y Aplicada (INFINA), Ciudad Universitaria, 1428 Buenos Aires, Argentina}
\author{M.~Noseda}
\affiliation{Universidad de Buenos Aires, Facultad de Ciencias Exactas y Naturales, Departamento de Física, Ciudad Universitaria, 1428 Buenos Aires, Argentina}
\affiliation{CONICET - Universidad de Buenos Aires, Instituto de F\'{\i}sica Interdisciplinaria y Aplicada (INFINA), Ciudad Universitaria, 1428 Buenos Aires, Argentina}
\author{P.J.~Cobelli}
\affiliation{Universidad de Buenos Aires, Facultad de Ciencias Exactas y Naturales, Departamento de Física, Ciudad Universitaria, 1428 Buenos Aires, Argentina}
\affiliation{CONICET - Universidad de Buenos Aires, Instituto de F\'{\i}sica Interdisciplinaria y Aplicada (INFINA), Ciudad Universitaria, 1428 Buenos Aires, Argentina}
\author{P.D.~Mininni}
\affiliation{Universidad de Buenos Aires, Facultad de Ciencias Exactas y Naturales, Departamento de Física, Ciudad Universitaria, 1428 Buenos Aires, Argentina}
\affiliation{CONICET - Universidad de Buenos Aires, Instituto de F\'{\i}sica Interdisciplinaria y Aplicada (INFINA), Ciudad Universitaria, 1428 Buenos Aires, Argentina}

\begin{abstract}
We combine experiments in a von K\'arm\'an flow with numerical simulations of Taylor-Green and homogeneous and isotropic turbulence to study the effect of the local flow geometry on particle pair dispersion. To characterize particle dispersion we use the pair dispersion angle, defined as the angle between the relative position and relative velocity of particle pairs. This angle was recently introduced as a means to more effectively identify the different dispersion regimes in finite-Reynolds-number flows. Our results show that, at a global scale, all flows considered show similar dispersion properties in terms of this metric, characterized by ballistic, super-diffusive, and diffusive regimes. Locally, however, these systems exhibit distinct behaviors, with anisotropies and local geometric features significantly influencing dispersion in both
the von K\'arm\'an and Taylor-Green flows.
\end{abstract}

\maketitle

\section{\label{sec:level1} Introduction}

The mixing of particles in turbulent flows is a significant and contemporary challenge with wide-ranging applications in both environmental and industrial contexts \cite{Pumir_2000, Falkovich_2001, Mordant_2004, Rast2011, Thalabard_2014, Rast_2016}. Advancing our understanding in this field is essential for several key areas. In climate modeling it is crucial for the process of droplet formation, which influences precipitation efficiency and the optical properties of clouds \cite{Kostinski_2005, Falkovich_2002, Shaw_2003, Ichihara_2023}. Accurately determining aerosol particle deposition is vital, as an example, during nuclear accidents involving radioactive particles \cite{Brooke_1992}. Finally, the efficiency of various industrial processes relies heavily on the effectiveness of turbulence as a mixing mechanism \cite{Oldshue_1983, Paul_2004}.

Both practical applications and fundamental turbulence research, specially when considering environmental flows, require consideration of turbulent flows that deviate from the conventional assumptions of homogeneity, isotropy, or stationarity, adding further complexity and significance to this problem \cite{Zapata_2024}. This is clearly the case in natural flows, but also in laboratory experiments as even in a carefully designed setup the flow invariably retains some information of the geometry of the enclosure or some memory of the forcing mechanism, and thus is never perfectly homogeneous or isotropic \cite{Biferale2005b, Shen2000}. Therefore, many studies in recent years considered the effects of anisotropy on different statistical properties of Lagrangian and of inertial particles \cite{BURRY1993, Pitton2012, Polanco2018, Angriman_2022_3, Shnapp_2024}.   

Among the various methods to study particle transport and dispersion, two approaches have traditionally garnered the most interest. While single particle statistics can effectively capture average particle transport, a more comprehensive understanding is achieved through the study of particle pair statistics. This approach is particularly important because theoretical descriptions of turbulence often rely on the properties of particle pairs \cite{Falkovich_2001, Sawford_2001, Salazar_2009}. For example, understanding the statistical properties of pair dispersion is crucial for computing correlations of passive scalars transported by turbulent flows \cite{Boffetta_2000}. 

The pioneering work in the study of particle pair dispersion in turbulence was conducted by Richardson \cite{Richardson1926}. Since then, in homogeneous and isotropic turbulence (HIT) three distinct regimes have been identified for particles with starting separation $\Delta r_0$ within the flow's inertial range, $\eta \ll \Delta r_0 \ll L_0$, where $L_0$ is the flow integral scale and $\eta$ is the Kolmogorov scale. Initially particles move with almost constant velocity, and their mean squared displacement thus follows a ballistic law, $\langle |\vb{r}_2 - \vb{r}_1|^2 \rangle = \langle |\Delta \vb{r}|^2 \rangle \sim t^2$. This regime ends at a characteristic time called the Batchelor time, $t_b = (\Delta r_0^2/\epsilon)^{1/3}$, which is the time needed for the particles to forget their initial condition, and where $\epsilon$ is the energy dissipation rate. Note $t_b$ corresponds to the correlation time of the eddy at scale $\Delta r_0$, which initially has the strongest influence on the particles' relative motion. Once particles forget the correlation imposed by their initial condition, for $t_b \ll t \ll T_0$ where $T_0$ is the flow integral turnover time, the effect of turbulent structures results in superdiffusive behavior with $\langle |\Delta \vb{r}|^2 \rangle \sim t^3$. Finally, for distances larger than the flow integral scale (or for times larger than $T_0$), the correlation imposed on the particles by the flow structures is lost, and the average behavior is expected to become diffusive, $\langle |\Delta \vb{r}|^2 \rangle \sim t$, as in Brownian motion \cite{Bourgoin2006}.

The superdiffusive behavior between times $t_b$ and $T_0$ results in the power law for pair particle dispersion known as Richardson's law, which is considered one of the hallmarks of turbulence. However, validating this power law experimentally or numerically  presents significant challenges. Among other factors, finite Reynolds number effects cause the statistics of particle separation to depend critically on their initial conditions \cite{Biferale2005}, complicating the theoretical contrast and analysis. In laboratory experiments the problem also requires having enough information of long particles' trajectories to get a statistically meaningful average to identify the power law, but also from particles starting initially from very small separations so that $t_b \ll T_0$ \cite{Bourgoin2006}. Another factor to consider is the effect of intermittency, which imposes long time correlations that affect the scaling observed in pair dispersion \cite{Boffetta2002}. 

These difficulties underscore the importance of continuing research in turbulence to better understand particle dispersion. To overcome some of them authors have come up with different solutions, such as the introduction of specific times to consider in the scaling (as, e.g., delay and exit times \cite{Bourgoin2006, Rast2011}), or the analysis of idealized systems \cite{Rast2011, Rast_2016}. A different approach was introduced recently in \cite{Shnapp2023}. In this study the authors considered the statistics of the instantaneous angle between the particles' relative position, $\Delta \vb{r}$, and their relative velocity, $\Delta \vb{v} = \Delta \vb{\dot{r}}$, defined as
\begin{equation}
    \cos(\theta) = {\Delta \vb{r} \cdot \Delta \vb{v} \over \abs{\Delta \vb{r}}\abs{\Delta \vb{v}}}
    = {1\over \abs{\Delta \vb{v}}} \dv{t} \abs{\Delta \vb{r}} .
    \label{eq:angle}
\end{equation}
This angle quantifies how aligned the dispersion process of particle pairs is. In \cite{Shnapp2023} it was shown that the average pair dispersion angle (APDA), $\langle \theta \rangle$, behaves differently in the three regimes mentioned above. When selecting random particle pairs with a given initial separation, velocity differences and separation vectors have random directions, and the statistical expectation is that $\langle \theta \rangle = 90 \degree$, as there is no preferential alignment. As particles evolve, for times $t \ll t_b$ corresponding to the ballistic regime, the APDA decreases linearly with time, $\langle \theta \rangle = \pi [ 1 - (t/t_b)(2 C_{1, \perp}/\pi)]/2$, where $C_{1, \perp}$ is the prefactor in the first-order transverse absolute structure function of the velocity. This results from the fact that particles in this regime move with approximately constant velocity and, on average, away from each other, which makes the APDA tend to zero. Once particles forget their initial condition, the behavior of the APDA should not depend on $\Delta r_0$ as the flow is self-similar, at least for separations smaller than $L_0$. As the only relevant physical magnitudes for the particles' pairs in the inertial range are $\epsilon$ and $t$, there is no dimensionless number that can be constructed from then, and thus $\langle \theta \rangle$ should remain constant. Numerical and experimental evidence presented in \cite{Shnapp2023} indicate $\langle \theta \rangle = (59 \pm 2) \degree$, which the authors claim is universal for homogeneous and isotropic turbulence in the  inertial range. Finally, for $t \gg T_0$, as trajectories decorrelate, the APDA starts increasing towards $90 \degree$ at a (presumably) flow-dependent rate.

In this work we study the behavior of the APDA in turbulent flows that deviate from isotropy and homogeneity. Specifically, we analyze the APDA for Lagrangian tracers across three distinct types of flows. First, we consider a von Kármán (VK) swirling flow experiment, where two counter-rotating propellers generate a turbulent flow with a large-scale circulation and with large Reynolds number \cite{VOLK_2011, Mordant_2004, Angriman2020}. Next, we perform and analyse results from direct numerical simulations (DNSs) of flows with Taylor-Green (TG) forcing, which generates two large-scale counter-rotating vortices that share similarities with the VK flow, despite differences in the forcing mechanism and the boundary conditions \cite{Mininni_2014, Angriman2020, Angriman_2021}. Finally, we examine DNSs with random forcing to generate HIT. The APDA can identify the different regimes of particle pair dispersion in all these flows, as well as quantify the impact of the local flow geometry in the dispersion process. In particular, we show that long-time lived saddle points in the VK and TG flow significantly affect the APDA and dispersion, using numerical and experimental data as well as theoretical models of the large-scale circulation in each of these flows.

\section{\label{sec:level2} Description of the datasets}

The VK experimental setup consists of a tank of $(20 \times 20 \times 50)$ cm$^3$, with two facing disks of diameter $D = 19 \ \rm{cm}$ separated by a vertical distance of $H = 20 \ \rm{cm}$. Each disk is fitted with eight radially oriented straight blades, each having a height and width of 1 cm; limited in length so they do not intersect at the center (see Fig.~1 and \cite{Angriman2020}). These two disks with blades work as impellers, driven by two independent brushless rotary motors (Yaskawa SMGV-20D3A61, 1.8 kW) controlled by servo-controllers (Yaskawa SGDV-8R4D01A). A forcing frequency of $f_0 = 1/T_0 = 50 \ \rm{rpm}$ was used in all experiments, with the two impellers counter-rotating. This generates two counter-rotating structures in the working fluid, two toroidal secondary circulations, and a strong shear layer on the midplane of the experiment \cite{Berning2023}. The generated flow is anisotropic, as is the forcing mechanism, which results in large-scale structures parallel to the disks' planes (horizontal) that have a larger correlation length than their axial (vertical) counterparts. Vertical and horizontal velocities are thus anisotropic \cite{Angriman_2021}. On the back of the impellers, two coils connected to a chiller allow heat removal, keeping the temperature of the working fluid at 25 $\degree$C. We used distilled water from a double-pass reverse osmosis system to remove ions and any dissolved or suspended solid particles.


\begin{figure}
    \centering
    \begin{subfigure}{0.35\textwidth}
      \centering
      \begin{picture}(0,0)
        \put(-100,-10){(a)}
      \end{picture}
      \includegraphics[width=1\linewidth]{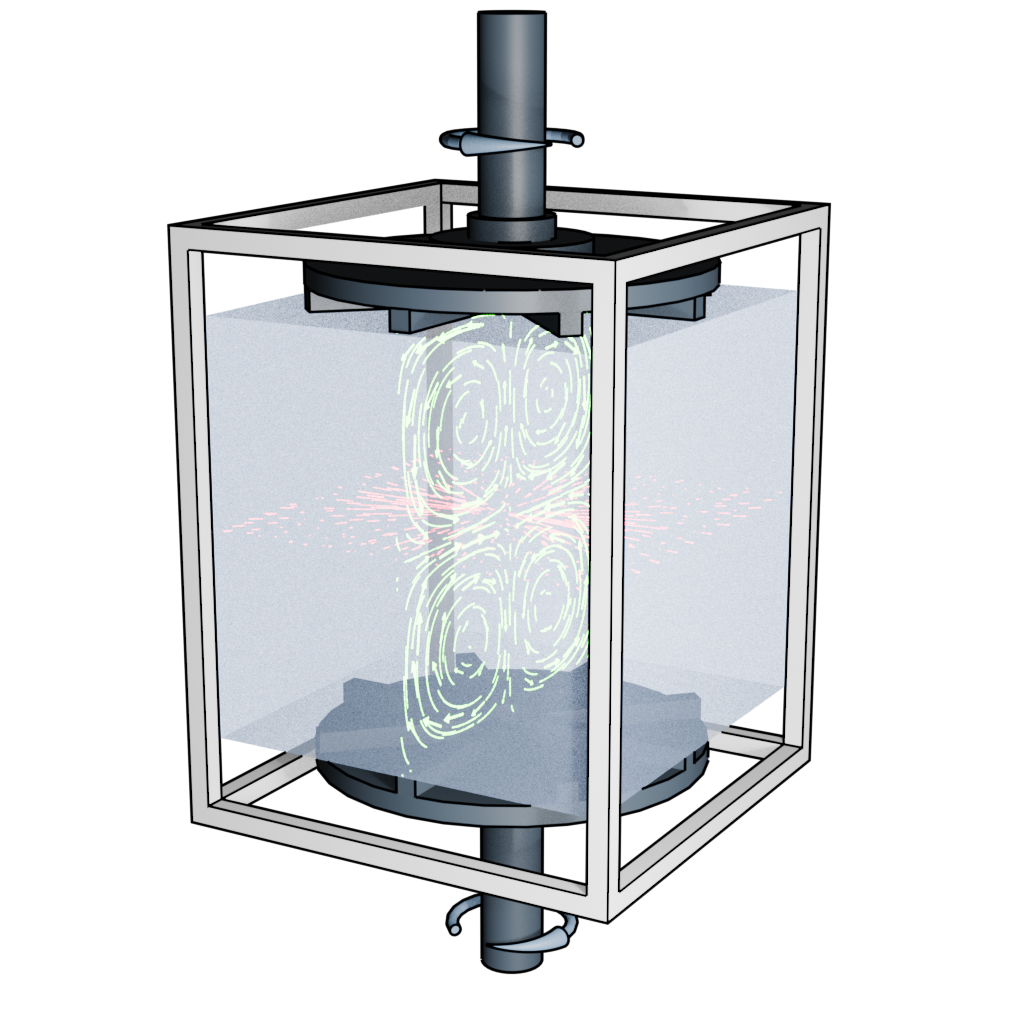}
    \end{subfigure}
    \hspace{2.5cm}
    \begin{subfigure}{0.35\textwidth}
      \centering
      \includegraphics[width=1.0\linewidth]{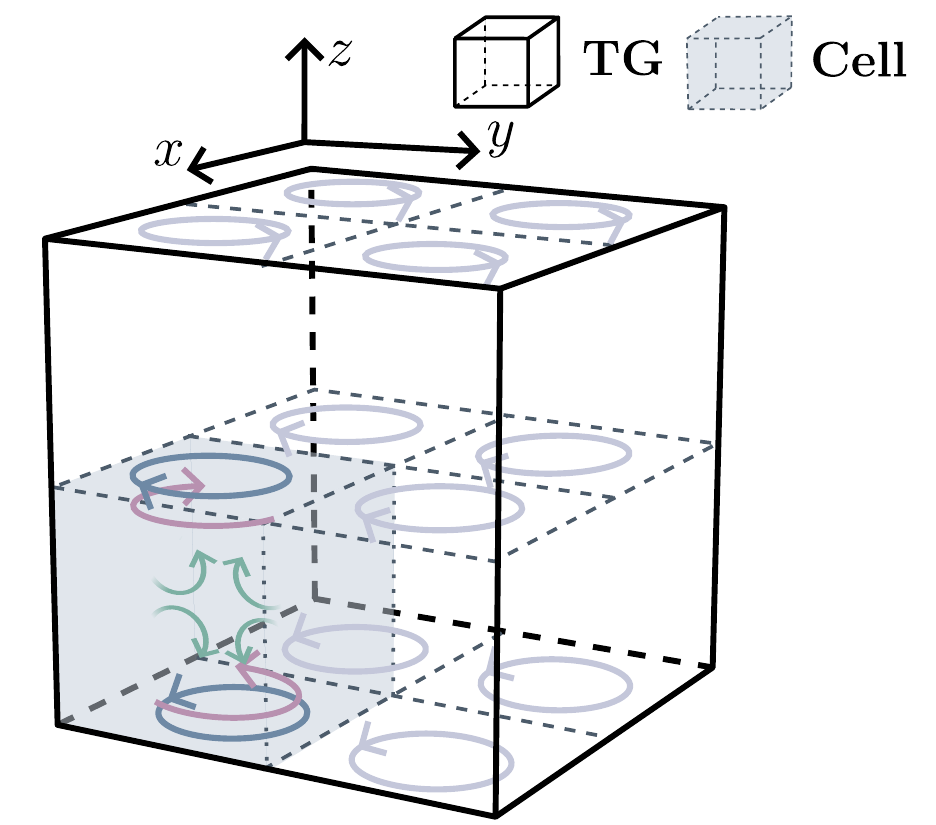}
			\vspace{.1cm}
      \begin{picture}(0,0)
        \put(-100,166){(b)}
      \end{picture}
    \end{subfigure}
    \caption{
        (a) Experimental setup. $H$ is the separation between the two impellers, $D$ is the diameter of the impellers, $h$ is the horizontal length of the cell, and $f_0$ denotes the rotation frequency of the impellers. The arrows inside de cell illustrate the principal features of the mean large-scale flow.
        (b) Diagram of the Taylor-Green simulations.
        The $(2\pi L_0)^3$ periodic domain is outlined with solid black lines, and $(\pi L_0)^3$ cells with large-scale structures reminiscent of the VK flow are indicated with dotted lines.}
    \label{tg_scheme}
\end{figure}
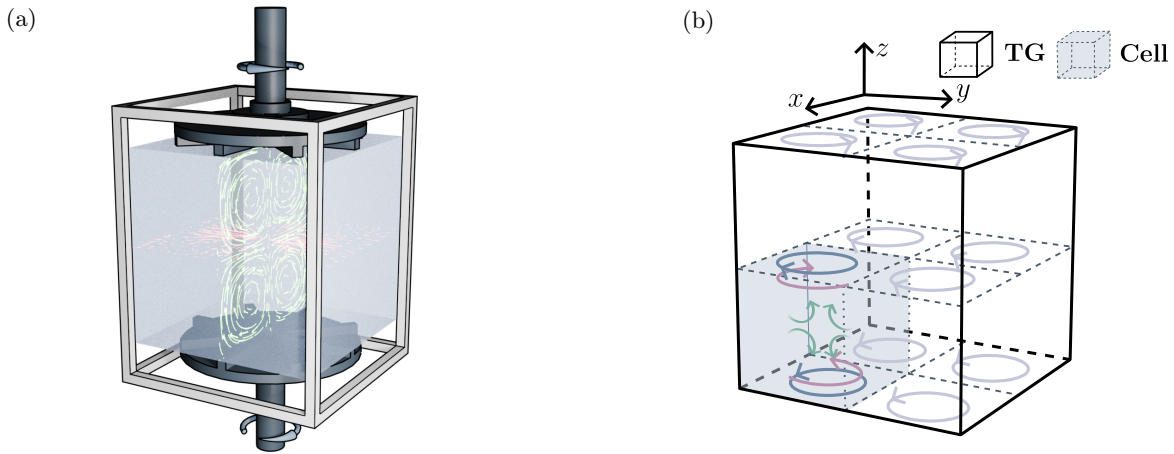

We seeded the flow with neutrally buoyant polyethylene microspheres (density $1 \, \rm{g} \, \rm{cm}^{-3}$) of diameter $d = 250-300 \ \mu\rm{m}$ (Cospheric), which we used as tracers. Particles are illuminated from two adjacent sides using two $25 \times 25 \ \rm{cm}^2$  LED panels (each 1880 lm, 22 W). We performed three-dimensional Particle Tracking Velocimetry (3D-PTV) using two high-speed Photron FASTCAM SA3 cameras with a resolution of $(1024 \times 1024) \ \rm{px}^2$ and 12-bit color depth at a sampling frequency of $f_s = 1000 \ \rm{Hz}$, in order to have close to 4 images per Kolmogorov timescale, $\tau_\eta = (\nu/\epsilon)^{1/2} = 0.011 \ \rm{s}$, where $\nu$ is the kinematic viscosity of water and $\epsilon$ is the average energy dissipation rate. The position of the cameras allows for tracking of the particles in a centered region of the setup with $(16 \times 16 \times 16) \ \rm{cm}^3$ volume. Additional details regarding the experimental setup can be found in \cite{Angriman2020, Angriman2022}. We performed twenty realizations of the experiment, reconstructing $\mathcal{O}(10^4$) particle trajectories with a mean duration of $0.3 \, T_0$.

The numerical datasets stem from DNSs of particle laden flows. For the TG dataset we performed a DNS of the incompressible Navier-Stokes equation using an external forcing $\mathbf{F}$ to sustain a turbulent regime. The equations are solved in a 3D cubic, $(2 \pi L_0)^3$-periodic domain where $L_0$ is a unit length, using a parallel pseudospectral method with the GHOST code \cite{Mininni2011, Rosenberg2020}, with a spatial resolution of $N^3$ = $768^3$ grid points. The forcing $\mathbf{F}$ is given by
\begin{equation}
    F_x = F_0 \sin(k_F x) \cos(k_F y) \cos(k_F z), \ \ \ F_y = -F_0 \cos(k_F x) \sin(k_F y) \cos(k_F z), \ \ \ F_z = 0, 
\end{equation}
with forcing wave number $k_F = 1$. This forcing results in two counter-rotating large-scale vortices in a $(\pi L_0)^3$ subvolume, with a secondary poloidal circulation. The similarities and differences between these two flows have been thoroughly studied (see, e.g., \cite{Angriman2020}). The eight subvolumes of size $(\pi L_0)^3$ inside the entire $(2 \pi L_0)^3$ volume will be considered separately to increase the statistics. These regions are visualized schematically in Fig.~\ref{tg_scheme}.

A HIT simulation was carried out following the same approach as in the TG DNS, with a spatial resolution of $768^3$ grid points. To sustain the turbulent regime, a forcing function was applied of the form
\begin{equation}
{\bf F}_\text{HIT} = F_1 \sum_{\abs{\vb{k}}\in(0, 2]} \text{Re}
\left\{
    \frac{i \vb{k} \times \vu{e}}{\abs{\vb{k}}}
    e^{i\,(\vb k \cdot \vb r + \phi_{\vb k})}
\right\},
\end{equation}
where $F_1$ is a constant amplitude, Re denotes the real part, $\vu{e}$ is some unit vector, ${\bf k}$ is the wave vector, and $\phi_{\vb k}$ is a random phase. The random phases for each mode evolve slowly over time, with a correlation time of 0.5 large-scale eddy turnover times.

The amplitudes $F_0$ and $F_1$ of the forcings in the TG and HIT simulations were adjusted to have an r.m.s.~velocity of order one, and the kinematic viscosity $\nu$ was adjusted to have well resolved simulations with $\kappa \eta > 1$ where $\kappa=N/3$ is the maximum resolved wave number. When the turbulent steady state in these simulations was reached, we evolved in time the position and velocity of $10^6$ Lagrangian particles according to $d \mathbf{x}_P/dt = \mathbf{u}(\textbf{x}_P, t)$, where $\mathbf{x}_P$ is the particle position at time $t$, and $\mathbf{u}(\mathbf{x}_P, t)$ is the velocity of the fluid element at position $\mathbf{x}_P$.

The Taylor-scale Reynolds number in the experiment and simulations is defined as
\begin{equation}
    \text{Re}_\lambda = \sqrt{15\, U^2 \over \nu \epsilon},
\end{equation}
where for the TG and VK flows $U$ is the root mean squared horizontal velocity of the particles, 
\begin{equation}
    U = \sqrt{\langle u_x^2 + u_y^2\rangle/2}. 
\end{equation}
while in HIT $U=[\langle u_x^2 + u_y^2 + u_z^2\rangle/3]^{1/2}$. 
The resulting values for VK, TG, and HIT are respectively $\text{Re}_\lambda^\text{VK} = 383$, $\text{Re}_\lambda^\text{TG} = 351$, and $\text{Re}_\lambda^\text{HIT} = 443$. Finally, as in the two DNSs $U$ is $\mathcal{O}(1) U_0$ where $U_0$ is the unit velocity, and we are considering $L^3=(\pi L_0)^3$ cells, we define $f_0 = 1/T_0 = U/L \approx \pi^{-1} U_0/L_0$ for all numerical simulations.

For further details on the VK experiment, and on TG and HIT direct numerical simulations, see \cite{Angriman2020, Angriman2022, Angriman2022_2}.

\section{\label{sec:level3}Results}

\subsection{\label{sec:global}Global statistics}

\begin{figure}
  \centering
  \includegraphics[width=0.5\linewidth]{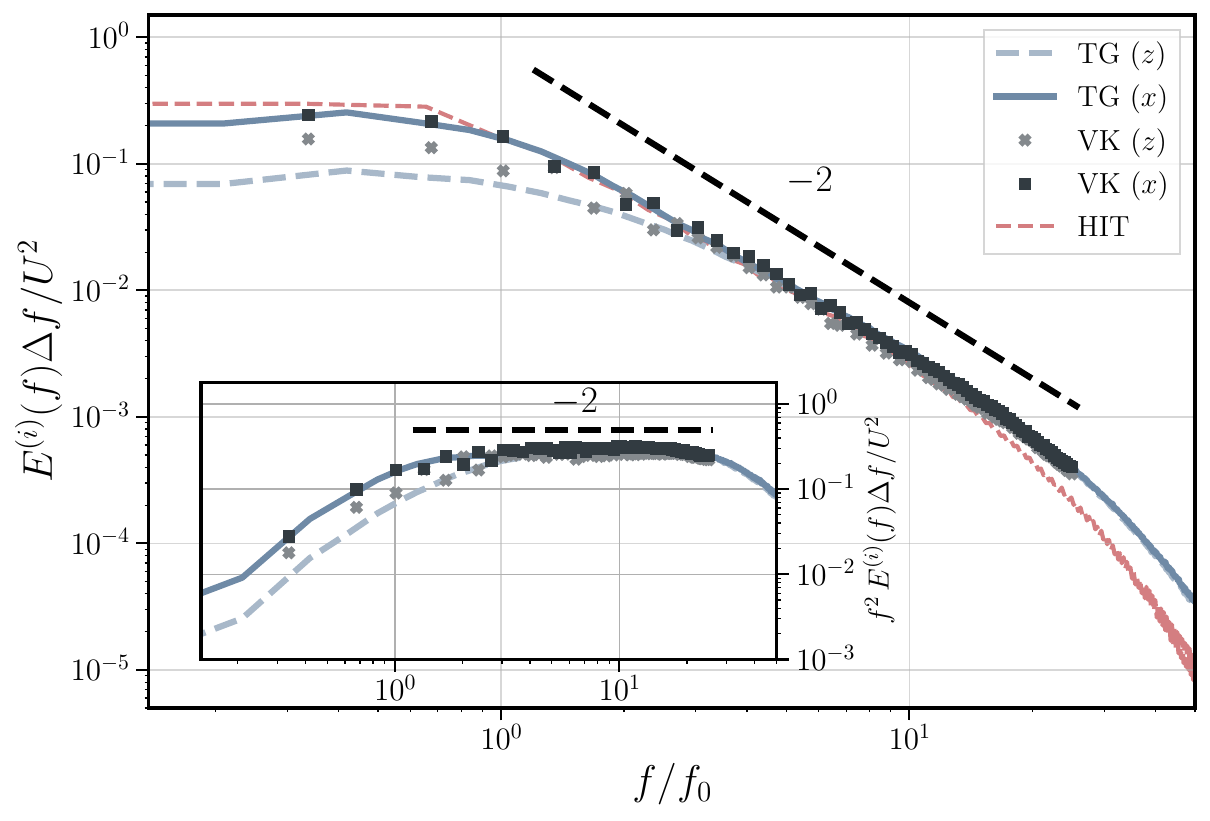}
  \caption{Lagrangian velocity power spectra for the $x$ and $z$ components of the tracers velocities in log-log scale for the VK, TG, and HIT datasets. The spectra are normalized by $U^2$ and the inverse of the frequency resolution $1/\Delta f$. A dashed line with $f^{-2}$ scaling is shown as a reference. The inset shows spectra compensated by $f^{-2}$.}
  \label{fig:spectrum}
\end{figure}

We start by characterizing particle dynamics with the Lagrangian velocity power spectrum obtained from the cosine transform of the autocorrelation function,
\begin{equation}
R_L^{i} = \frac{C_v(\tau)}{C_v(0)} = \frac{\langle v_i(t) v_i(t+\tau) \rangle}{\langle v_i ^2(t) \rangle},
\end{equation}
where the brackets $\langle \cdot \rangle$ denote an average over both time and trajectories, and $\tau$ represents the time lag. For the autocorrelation calculation in the experiment we only kept trajectories longer than $1/f_0$, as done in \cite{Angriman2020}. In Fig.~\ref{fig:spectrum} we show the averaged Lagrangian power spectrum of the HIT, TG, and VK datasets. We depict with a black dashed line and as a reference a $\sim f^{-2}$ scaling, the expected scaling for the Lagrangian velocity \cite{Toschi2009, Lvov1997} (see \cite{Angriman2020} for a discussion on corrections to this scaling from the mean flow in the VK and TG flows). Note that both the VK and TG datasets present a significant difference in the $x$ and $z$ spectra at low frequencies, resulting from the mean flow anisotropy. Except for this discrepancy, no other significant differences can be identified between the spectra of the different flows.

\begin{figure}
    \centering
    \includegraphics[width=.32\textwidth]{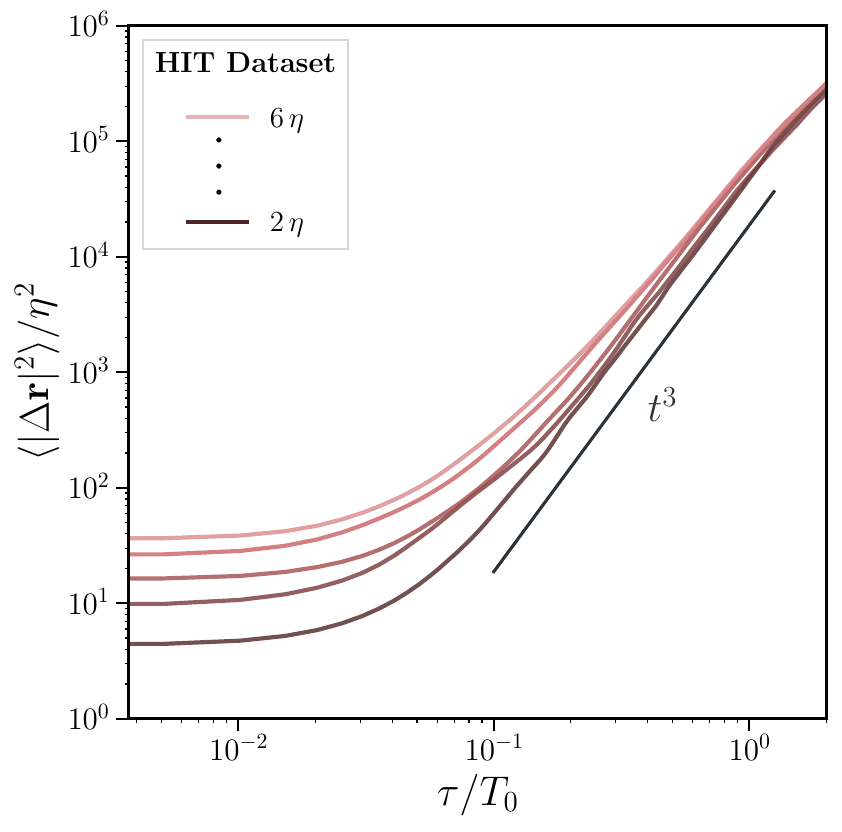}
    \includegraphics[width=.32\textwidth]{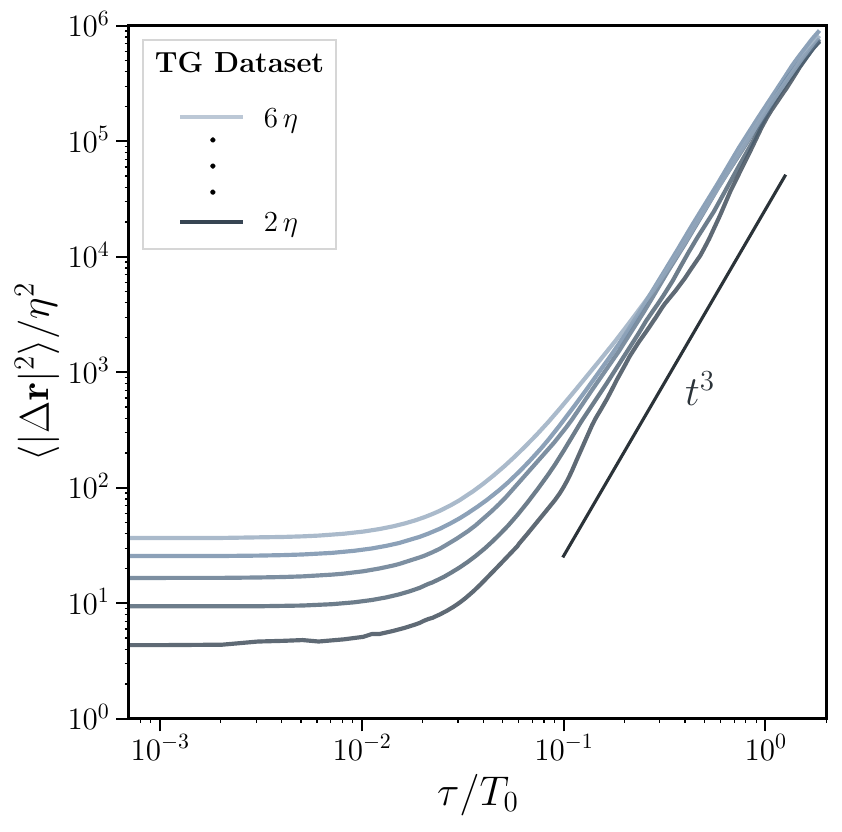}
    \includegraphics[width=.32\textwidth]{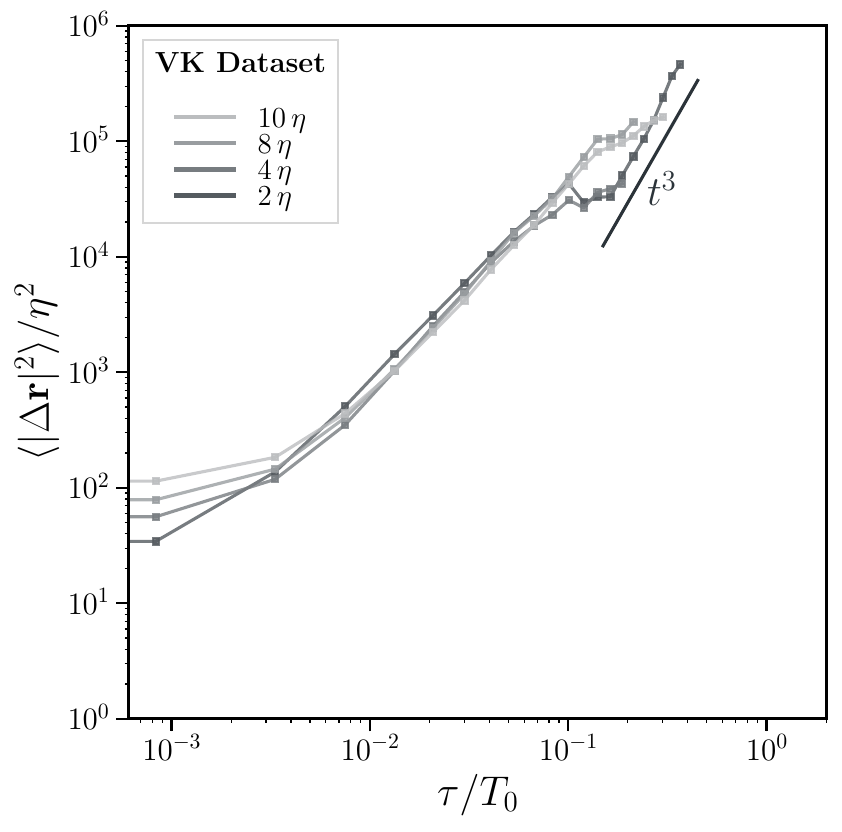}
    \caption{Mean squared distance between particle pairs normalized by the squared Kolmogorov dissipation scale, as a function of time, for different initial separations (see labels in the insets). Time is in units of the large-scale turnover time. From left to right, the panels show data for HIT, TG, and VK flows. A $\sim t^3$ scaling is shown as a reference.}
    \label{fig:msd}
\end{figure}

As previously mentioned, the usual way to characterize pair dispersion is to study the mean squared displacement of the particles. Figure \ref{fig:msd} shows this quantity computed for our three datasets, and for different initial separations of the particles' pairs. A Richardson's superdiffusive regime is visible in the HIT and TG simulations, using an ensemble of $\mathcal{O}(10^2)$ pairs of trajectories with initial separations $\Delta r_0 \in [2 , 6] \, \eta$, with bin width of $\eta$. For the VK experiment, given the smaller number of particles pairs available with initial separation in this range, we took a bin width of $4\eta$, and computed the mean squared dispersion for $\Delta r_0 \in \{2, 4,8, 10\}\,\eta$. Even with these choices, we are only left with $\mathcal{O}(10)$ pairs in the experiment for each bin. In simulations, the availability of longer time series permits the investigation of times beyond $t/T_0 \approx 1$, while experimental data presently support analysis up to around $t/T_0 \approx 5 \times 10^{-1}$.
Within this constraints, a limited range consistent with Richardson scaling is observed for particles with $\Delta r_0 = 2\eta$ and $4\eta$. The mean square dispersion also shows a dip before the superdiffusive range, around $\tau/T_0 = 10^{-1}$. This dip may result from limited statistics; however, we will later show that the APDA can identify the superdiffusive range and may offer insights into the origin of the deviations observed at $\tau/T_0 = 10^{-1}$.

\subsection{The average pair dispersion angle}

\begin{figure}
    \centering
    \includegraphics[width=.32\textwidth]{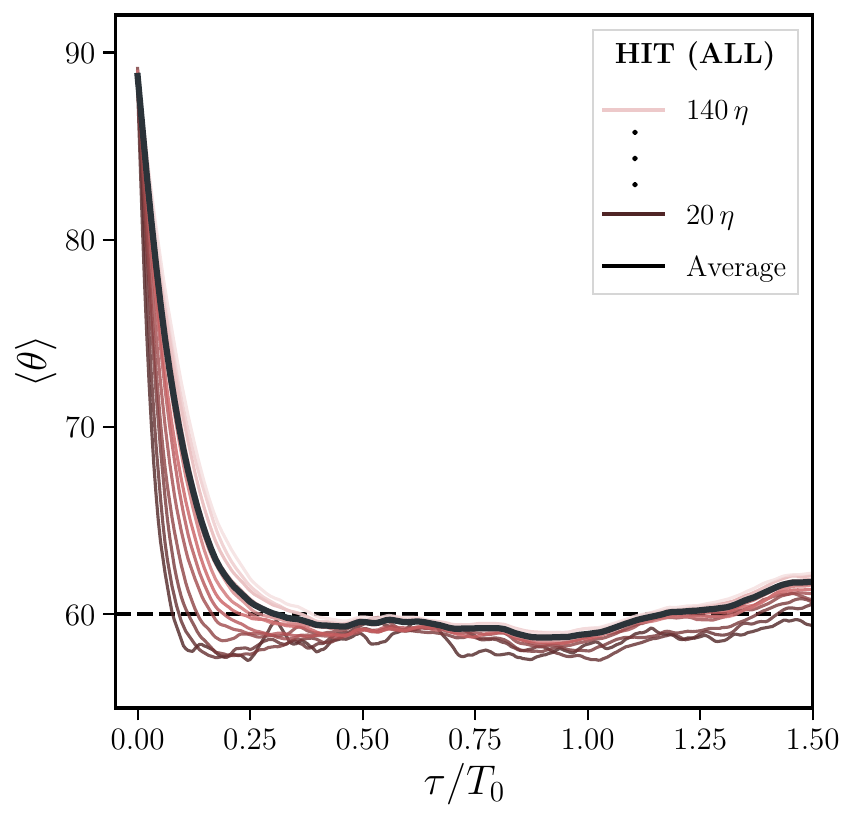}
    \includegraphics[width=.32\textwidth]{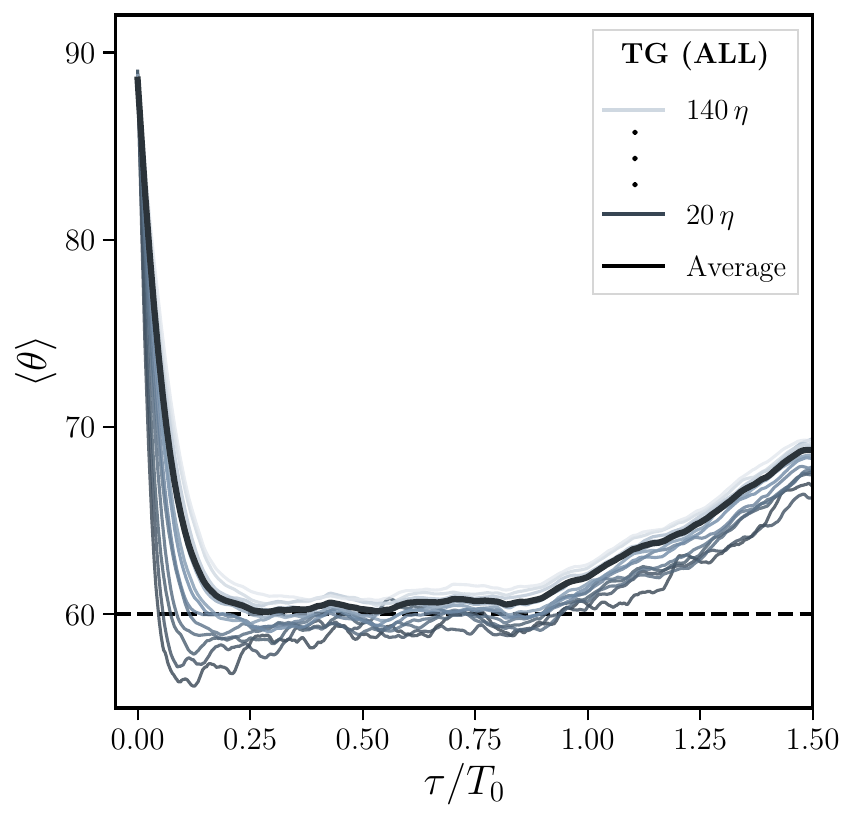}
    \includegraphics[width=.31\textwidth]{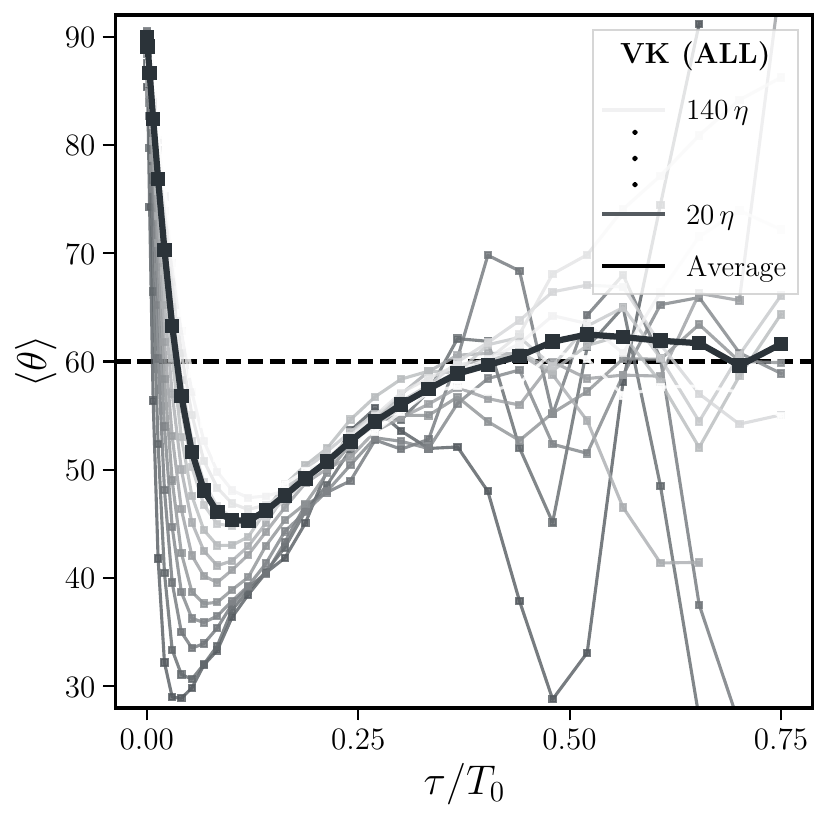}
    \caption{Average pair dispersion angle for the HIT (left), TG (center), and VK (right) datasets, calculated from particle pairs within entire cells. Curves in color show $\langle \theta \rangle$ for various initial separations, while thick black curves represent the overall average across all separations.}
    \label{av_pair_disp_angle}
\end{figure}

In this section, we focus on the analysis of the APDA across the three datasets previously described. Figure \ref{av_pair_disp_angle} shows the time evolution of this angle in the HIT, TG, and VK flows, for twelve different values of the particles' initial separations, in bins with mean initial distances $\Delta r_0 \in [20,140] \eta$ using bin widths of $10 \eta$. In all panels, the darkest color represents the smallest initial separation between pairs, and the brightest indicates the largest. The average $\langle \theta \rangle$ across all initial separations is shown in black.
For the DNS datasets, particle pairs were consistently selected within the same cell, as defined in Fig.~\ref{tg_scheme}, and averages were taken across the cells, treating them as eight independent flow realizations. Although this procedure does not affect the HIT dataset due to the homogeneity of turbulence in that case, we applied it nonetheless to ensure a consistent comparison across all DNS data. For the VK dataset, we employed a moving average on $\langle \theta \rangle$ to reduce fluctuations resulting from the smaller statistics available in the experiment.

The APDAs for all three datasets, as depicted in Fig.~\ref{tg_scheme}, begin at $90\degree$.
This is due to the fact that, at $t=0$, there is no preferential alignment in the separation of particles, as they are selected randomly. For $t > 0$ all datasets exhibit a sudden decrease in the APDA, corresponding to the onset of the ballistic regime. The datasets then reach a minimum of $\langle \theta \rangle$ at the end of the ballistic regime, for $t/T_0 \lesssim 0.25$, specially for particle pairs with smaller initial separations. In the VK dataset, as well as in the TG dataset for small initial separations, the presence of this minimum is particularly clear, with a smaller APDA compared to the HIT dataset. The reasons for this will be addressed in detail later. In all datasets, the APDA stabilizes around $t/T_0 \approx 0.25$, indicating the transition to the superdiffusive regime where $\langle \theta \rangle$ remains constant. Finally, in both DNS datasets, the APDA begins to rise at approximately $T/T_0 \approx 1$, marking the onset of the diffusive regime. This regime is not accessible in the VK experiment, as particles typically do not remain within the observation region for sufficiently long times.

Since these curves are derived using a first-order approximation of the average angle, we verified that the Taylor error in the computation remains constant and does not exceed 20\% across all cases considered, by replicating the analysis performed in \cite{Shnapp2023} for our datasets.

The APDA clearly distinguishes the three dispersion regimes, even when using initial spatial separations that are one hundred times greater than those chosen to compute the mean square displacement, along with bins that are ten times the size. This capability enables the characterization of pair dispersion processes in systems with fewer particles. For instance, in our experimental dataset, this metric allows for coarser segregation in the initial separation, making it particularly useful for analysis in less densely populated systems.

\begin{figure}
  \centering
  \includegraphics[width=0.4\linewidth]{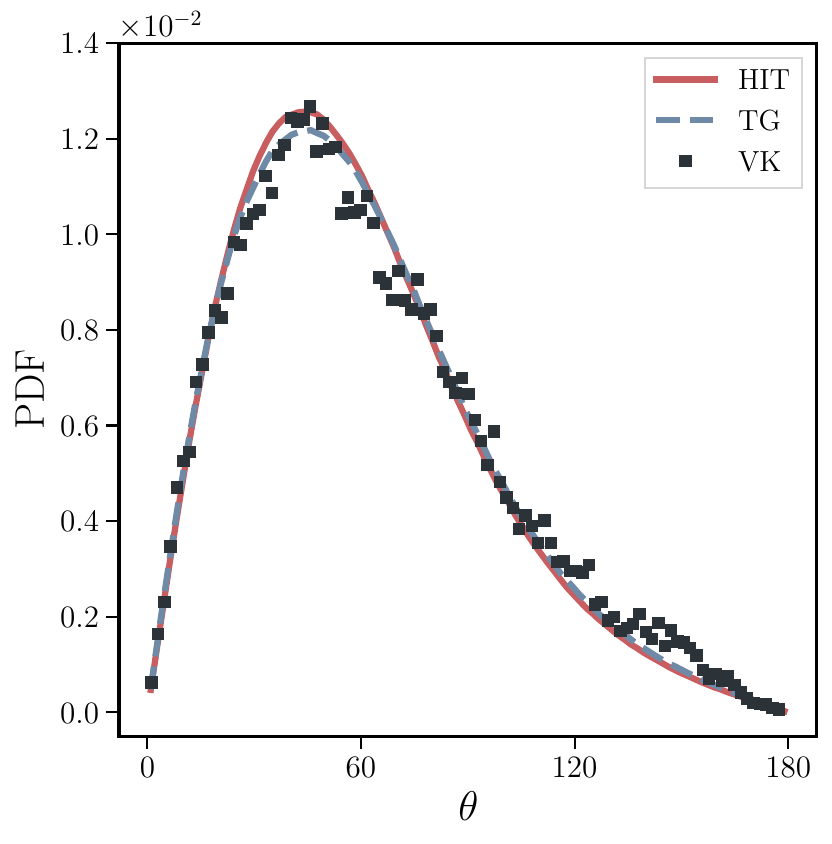}
  \caption{Probability density functions of the pair dispersion angle for all particle pairs within the plateau in Fig.~\ref{av_pair_disp_angle}, for the HIT, TG, and VK flows.}
  \label{pda_histogram}
\end{figure}

To compare the average values attained during the plateau across all datasets, we constructed the probability density function (PDF) for $\theta$ over the time intervals in which the APDA remains constant (see Fig.~\ref{pda_histogram}). The APDA for each dataset is $\langle \theta \rangle_{\rm{HIT}} \simeq 59.6 \degree$, $\langle \theta \rangle_{\rm{TG}} \simeq 60.4 \degree$, and $\langle \theta \rangle_{\rm{VK}} \simeq 62.2 \degree$.
The reported $\langle \theta \rangle$ in the simulations is an average over trajectories in the eight cells shown in Fig.~\ref{tg_scheme}, each cell having a slightly different mean value, with a standard error of $\sigma_{\langle\theta\rangle} = 2.3 \degree$ for both DNSs. These values are  statistically indistinguishable from one another and align closely with that reported in  \cite{Shnapp2023}. Despite differences in the flow characteristics of HIT, TG, and VK datasets, the PDFs of the dispersion angle in the inertial range exhibit a consistent statistical behavior. Remarkably, the anisotropies and inhomogeneities inherent to the TG and VK flows do not appear to have a significant effect on the statistics of $\theta$ within the superdiffusive regime.

Although these datasets share similarities, there are marked differences evident in Fig.~\ref{av_pair_disp_angle}. The most prominent one lies in the rate at which the APDA decreases from $90 \degree$ during the ballistic regime. In particular, the steep dip observed in the VK experiment, where $\langle \theta \rangle$ reaches a minimum around $\tau/T_0 \approx 0.1$, coincides with the transition in the dispersion seen in Fig.~\ref{fig:msd}. To understand the reasons for these discrepancies between the dispersion in the HIT, TG, and VK flows, and in particular for the strong decrease of the APDA at early times in some of these flows, we next look in detail at what distinguishes the three datasets: the local geometrical properties of their mean flows.

\subsection{\label{sec:local}Local statistics}

The VK and TG flows exhibit both anisotropy and inhomogeneity due to their forcing mechanisms, which create larg-scale mean flow structures characterized by distinct velocity stagnation points. Our hypothesis is that the presence of these zeros significantly influences particle transport and, consequently, the dispersion process. Stagnation points in turbulent flows have already been reported to affect transport, as, e.g., in the case of the clustering of inertial particles \cite{Mora_2021}. The local geometry and global topology of the flow is also known to affect trajectories \cite{Angriman_2021}, with inhomogeneous helicity also playing a role in local energy dissipation and transport \cite{Yokoi_1993, Yokoi_2016}. Consequently, in the following we conduct a detailed examination of regions with saddle points and their influence on the average pair dispersion angle.

As sketched in Fig.~\ref{tg_scheme}, both VK and TG flows display a saddle point in the velocity in the center of the cell (schematized by the color arrows). In practice, the position of this saddle point fluctuates in time, and is slightly different for the VK and TG flows. In the VK experiment, the forcing mechanism produces an Ekman pumping phenomenon in which pressure differences in the rotating propellers drive fluid from the center of the propellers to the periphery, leading, by mass conservation, to a displacement of fluid from the center of the cell towards the propellers. This process generates the poloidal flow marked by purple arrows in Fig.~\ref{tg_scheme}, and creates a stagnation point in the vicinity of the center of the cell.
In the TG flow, the secondary flow, marked by green arrows in Fig.~\ref{tg_scheme}, is generated instead by pressure differences caused by the volumetric forcing \cite{TaylorGreen1937}. The type of saddle formed then depends on the specific forcing and boundary conditions: in the TG case this point has two stable directions ($x$ and $y$) and one unstable direction ($z$) \cite{TaylorGreen1937}, while in the VK case there is one stable direction ($x$) and two unstable directions ($y$ and $z$) \cite{Berning2023}. Note that, as a result, perfect axisymmetry around the axis of the propellers is broken in the VK flow, and the choice of the $x$ and $y$ directions is arbitrary (see \cite{Berning2023} and more details in Sec.~\ref{subsec:low_order_models}).

Another interesting set of saddle points, generated by the primary flow in the TG forcing, is given by the points located at the intersection of the eight Taylor-Green cells that live in the full $(2\pi L_0)^3$ periodic domain; that is, the points in the corners of the TG cells.
These points are unique to the TG flow, and unlike the points that are in the center of the TG cells, they have one stable and two unstable directions. This makes them particularly interesting, as their configuration mirrors that of the saddle points at the center of the VK flow.
In addition to the previously mentioned stagnation points, another possible factor influencing particle transport in these flows is the intense shear layer in the (horizontal) midplane of each cell, created by velocity field gradients in the $z$ direction caused by the forcing. In the TG flow this gradient results from the modulation of the volumetric forcing in that direction, while in the VK flow it results from the counter-rotation of the propellers. The effect is especially dominant in the middle region of the cells.

In this light we analyze again the results shown in Fig.~\ref{av_pair_disp_angle}, this time  considering different subregions of the flow based on the aforementioned saddle points and regions with strong shear. We will see that while the results in Fig.~\ref{av_pair_disp_angle} may, at first glance, suggest that the differences in flow topology between TG, VK, and HIT have minimal consequences on the APDA, the apparent universality is actually caused by cancellations between the contributions of the different saddle points to the pair dispersion angle when averaging over the entire domain. In particular, we will show that the marked decrease in $\langle \theta \rangle$ at $\tau/T_0 \approx 0.1$ in the VK flow, relative to the TG and HIT datasets, can be linked to the presence and structure of its saddle points.

\begin{figure}
    \centering
    \includegraphics[width=\linewidth]{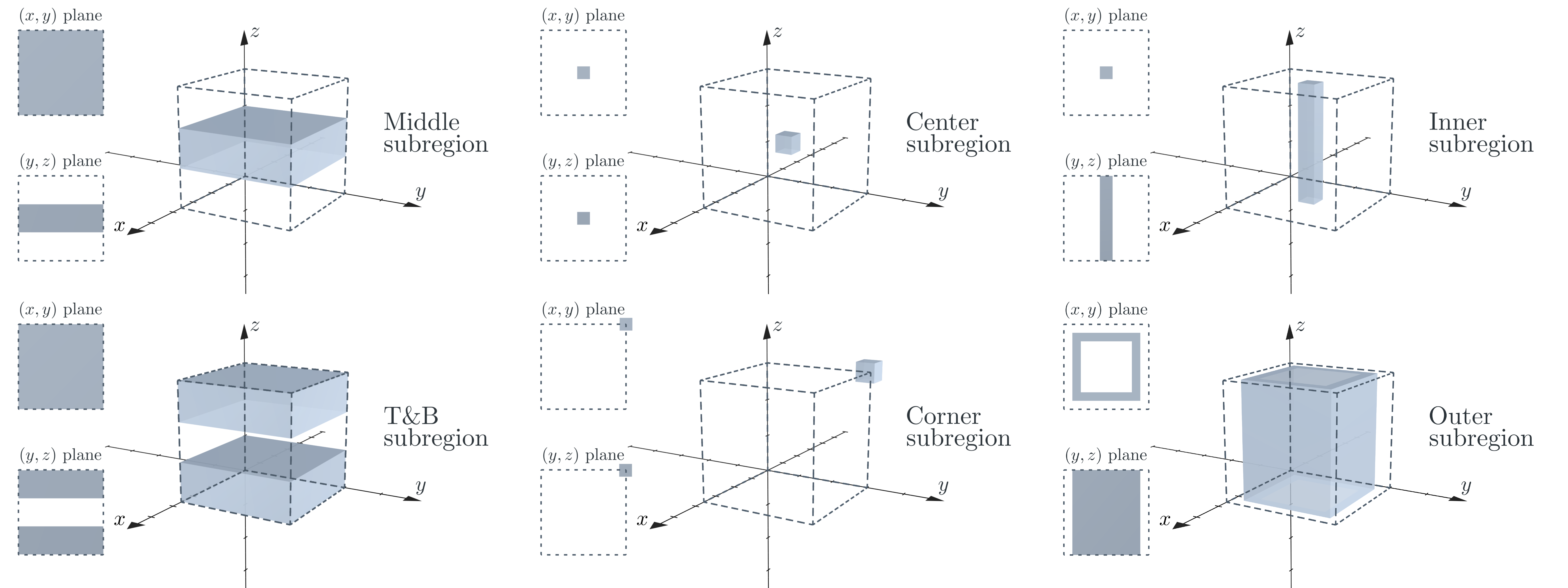}
    \caption{Different subregions considered in the analysis to compute local particle-pair statistics. The first separation, between middle and top-bottom (T\&B) subregions, sets apart the effect of the midplane shear layer from the rest of the flow. The second separation, in center and corner subregions, is intented to isolate the effect of saddle points in TG DNSs. Finally, the inner and outer subregions attempt the same separation between the central saddle point and its complement, while still preserving larger volumes and statistics of particles for the VK experiment.}
    \label{fig:subregs}
\end{figure}

The way we divide these flows into subregions is illustrated in Fig.~\ref{fig:subregs}. For the VK experiment, the sketch corresponds to the observation region. For the DNSs, the subregions are defined by subdividing the $(\pi L_0)^3$ cells rather than the entire $(2\pi L_0)^3$ periodic domain, to compare with the experiment on equal footing. The pair dispersion angle is therefore computed by considering only those trajectories of particle pairs initially belonging to each individual subregion.

In Fig.~\ref{fig:subregs} we begin by separating the middle subregion of the cell, where the dynamics are dominated by the strong shear layer, from the top and bottom subregions (labeled as T\&B), which contain strong large-scale vortices dominated by the forcing mechanism. Next, we individually isolate the vicinity of saddle points of the large-scale flow: the center subregion captures the stagnation points generated by the secondary flow, while the corner subregion (only relevant for the DNSs) captures the stagnation points generated by the primary TG flow.
Finally, we divide the flow into an inner subregion and a peripheral one (labeled as ``outer'' subregion). The inner subregion encloses again the saddle point in the center for the cell, plus the axis of rotation of the propellers, while the outer subregion constitutes its complement. This last separation allows us to capture the effect of the saddle point at the center of the VK, while also considering a larger subvolume. This approach will prove essential for gathering enough statistics and ensuring that the computation of the APDA is well-converged in the experiment.

\subsubsection{Middle and T\&B subregions}

\begin{figure}
    \centering
    \includegraphics[width=.32\textwidth]{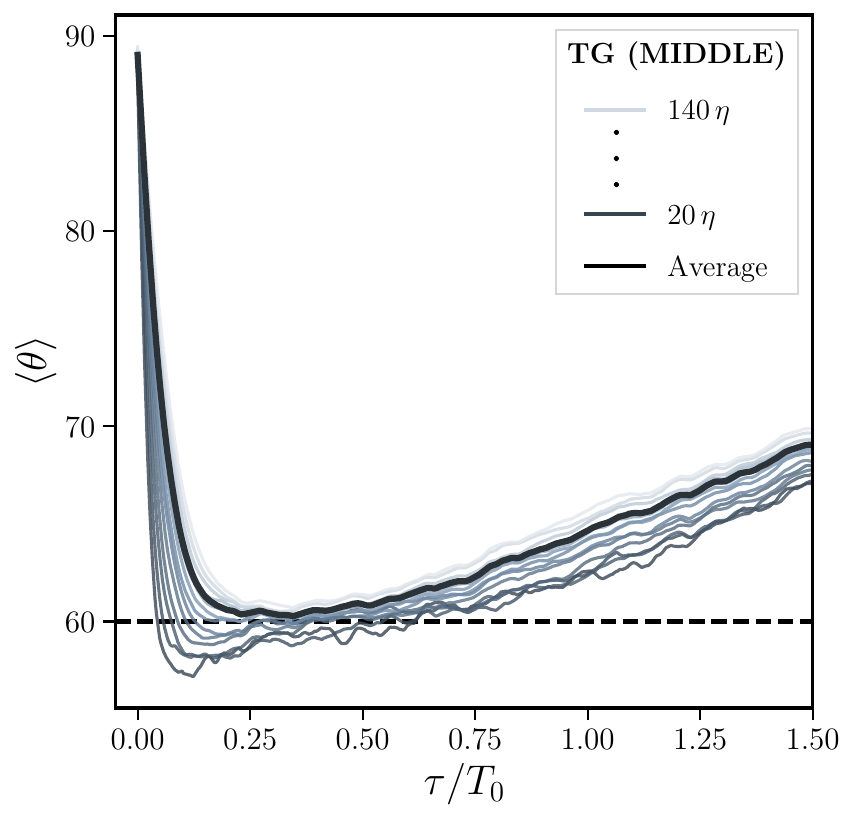}
    \includegraphics[width=.32\textwidth]{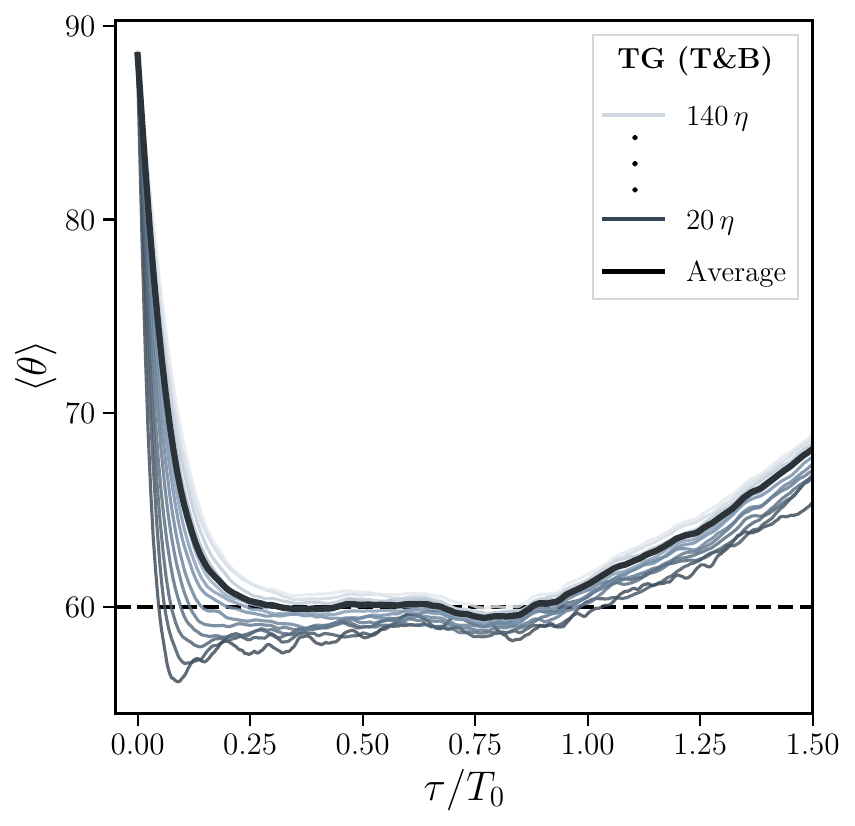}
    \includegraphics[width=.314\textwidth]{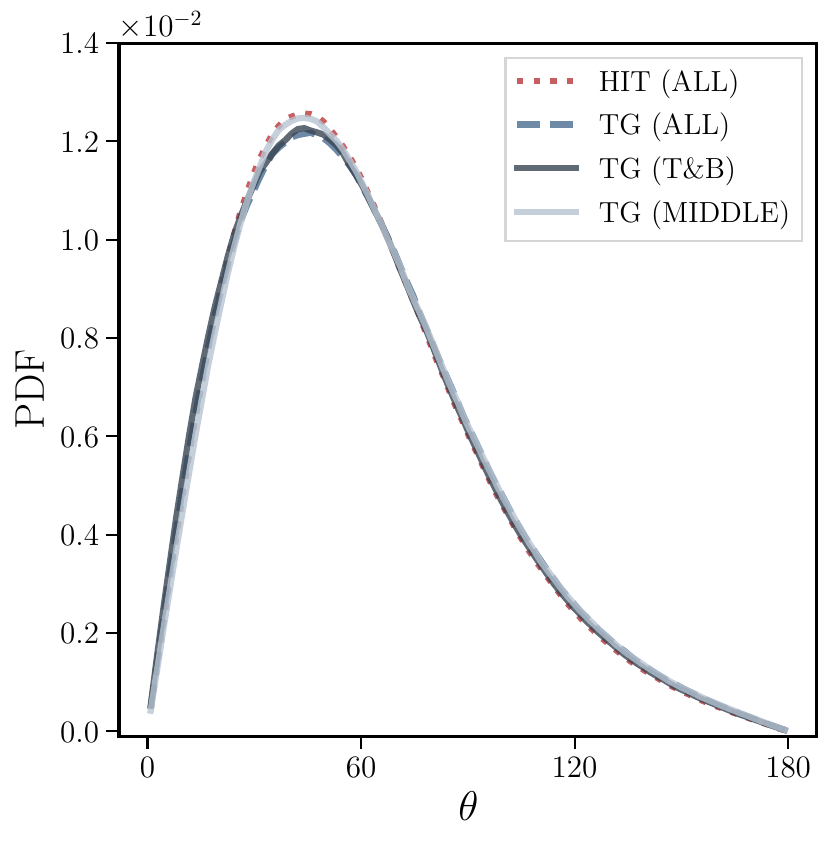}
    \caption{Average pair dispersion angle for the middle and T\&B subregions of the TG dataset, and corresponding PDFs of all the pair dispersion angles computed during the time window over which $\langle \theta \rangle$ remains approximately constant on average, for TG (in the entire cell, and in the middle and T\&B subregions) as well as for HIT simulations in the entire domain.}
    \label{fig:top_middle_subregion}
\end{figure}

In Fig.~\ref{fig:top_middle_subregion} we compare the APDA between the middle and T\&B subregions of the TG flow. The T\&B subregion does now show significant differences from the APDA in the entire cell (see Fig.~\ref{av_pair_disp_angle}, center panel). Even though this subregion has a strong and large-scale mean flow, its effect on the pairs of particles can be removed by a local Galilean transformation, leaving the relative velocity of the particles depending mostly on the turbulent fluctuations. Instead, the middle subregion displays a much shorter plateau (i.e., a shorter range of times in which the APDA is compatible with superdiffusive behavior). As this region is dominated by shear, this can be the result of a shorter local eddy turnover time, or the effect of the shear layer causing the directions at which particles move away from each other to decorrelate faster. In spite of these differences, if we compute the PDFs of the pair dispersion angle during the times in which the APDA remains approximately constat in each subregion, no significant differences between the PDFs and those from HIT are observed. Thus, the effect of the shear layer seems to be mostly associated with a change in the time scales of the system.

\subsubsection{Center and corner subregions}

\begin{figure}
    \centering
    \includegraphics[width=.32\linewidth]{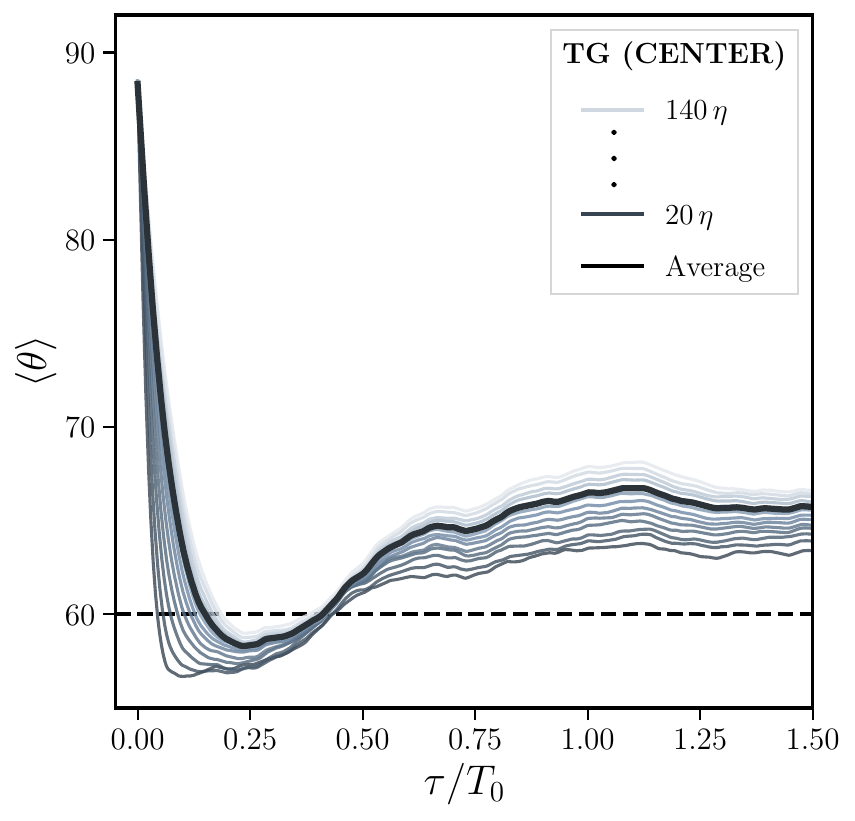}\hspace{1cm}
    \includegraphics[width=.32\linewidth]{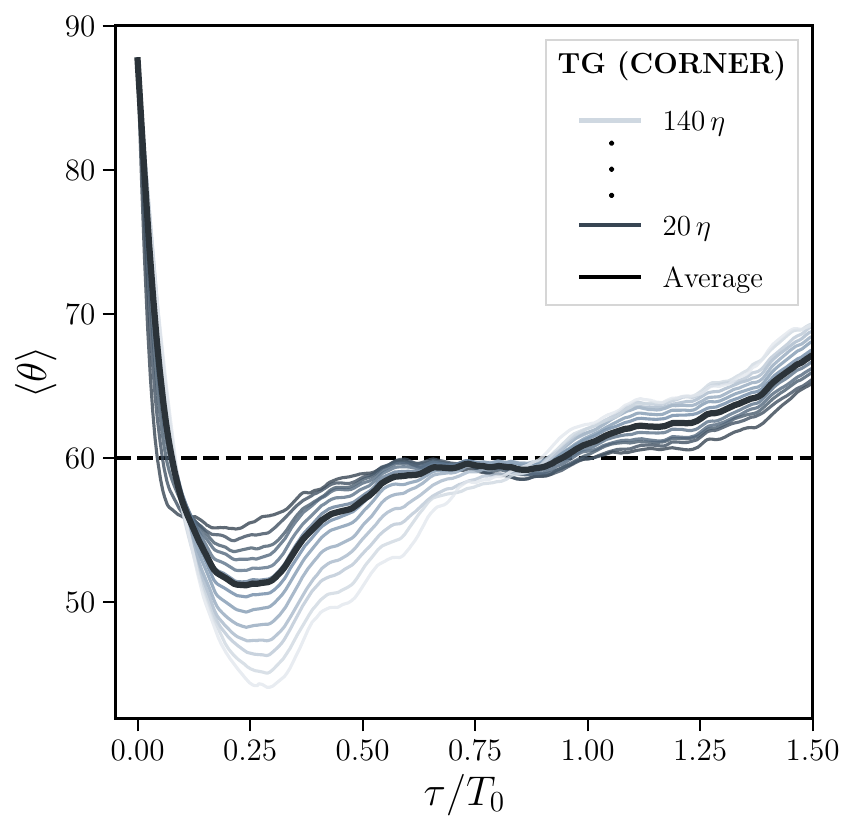}
    \caption{Average pair dispersion angle as a function of time for the center and corner subregions of the TG simulations.}
    \label{fig:tg_center_corner}
\end{figure}

The same analysis was carried out in these two subregions, to compare the properties of the saddle points in the TG flow.
Figure \ref{fig:tg_center_corner} shows that, in both the center and corner subregions, the APDA exhibits a sudden decrease at short times, followed by a recovery and a stabilization of $\langle \theta \rangle$ in the vicinity of some mean value which is larger for the center than for the corner subregion. This behavior, with the strong drop at early times, is reminiscent of that observed in the VK experiment (see Fig.~\ref{av_pair_disp_angle}, right panel).
The fast drop in the APDA indicates that at short times there is a strong preferential alignment between pairs of particles, driven by the saddle point dynamics.
Moreover, the two saddle points in the TG flow display another difference. While in the center subregion particles starting with a smaller initial separation $\Delta r_0$ have a faster drop and reach smaller minimal values of $\langle \theta \rangle$, the opposite happens in the corner subregion, with particles with a larger initial separation $\Delta r_0$ displaying the smaller minima in $\langle \theta \rangle$. Interestingly, the latter is also the behavior seen in the VK experiment in Fig.~\ref{av_pair_disp_angle}. We attribute the difference in these two subregions to the differences in the two saddle points in the TG cell, and in particular to the differences in their number of attracting and repelling directions. This points to the fact that the pair dispersion process near stagnation points can be strongly influenced by the saddle dynamics, providing a possible explanation for the dynamics observed in the VK flow. We explore this further using low-order models in Sec.~\ref{subsec:low_order_models}.

\subsubsection{Inner and outer subregions}

\begin{figure}
    \centering
    \includegraphics[width=.32\textwidth]{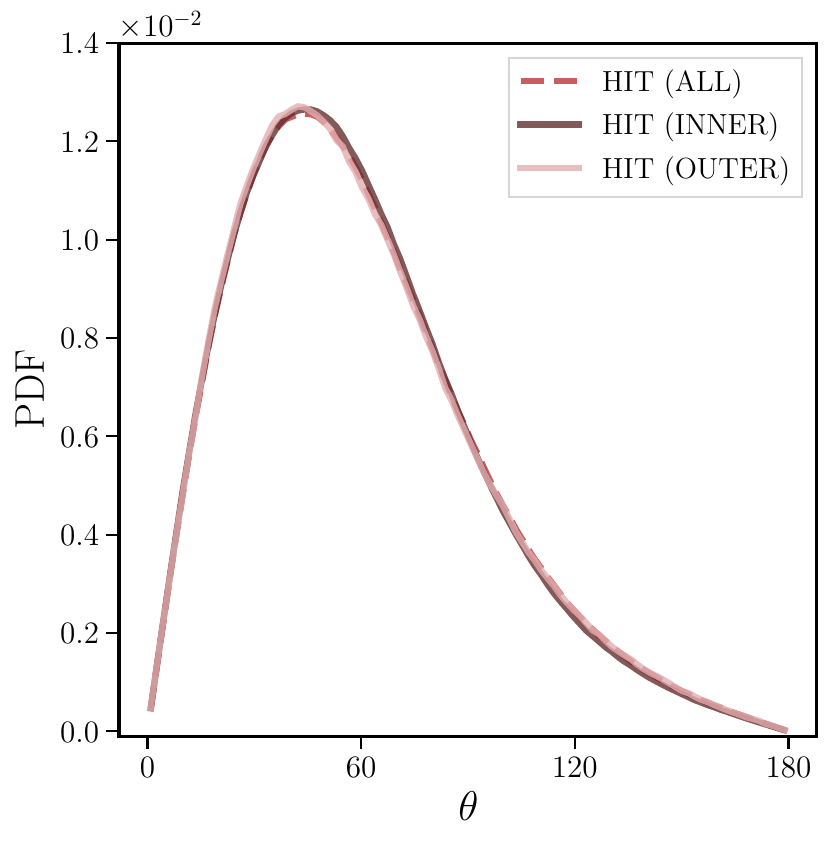}
    \includegraphics[width=.32\textwidth]{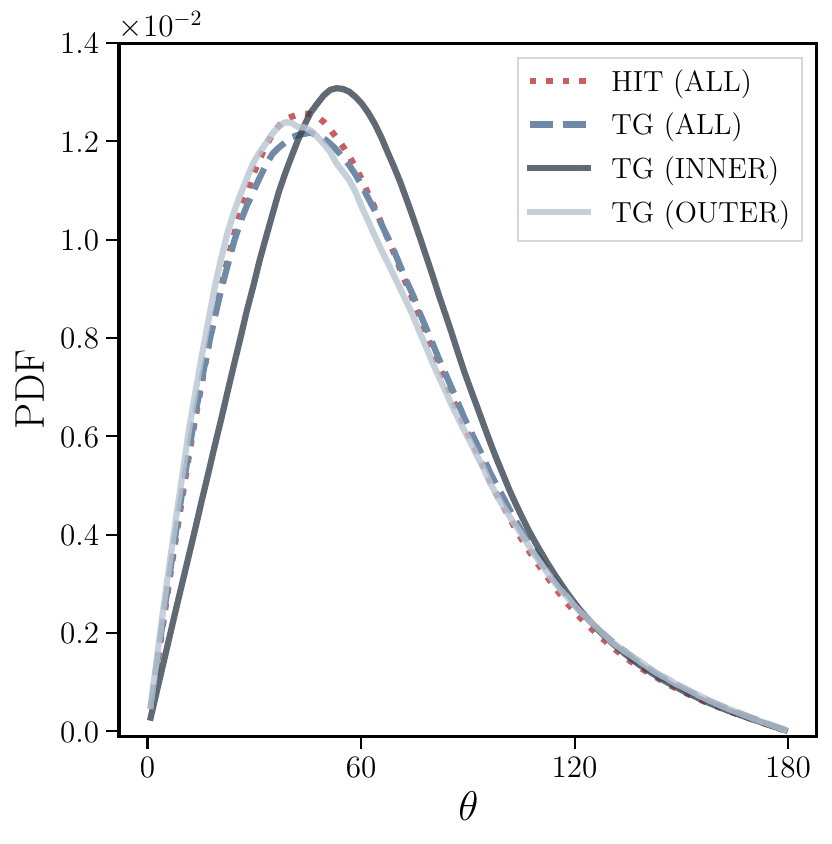}
    \includegraphics[width=.32\textwidth]{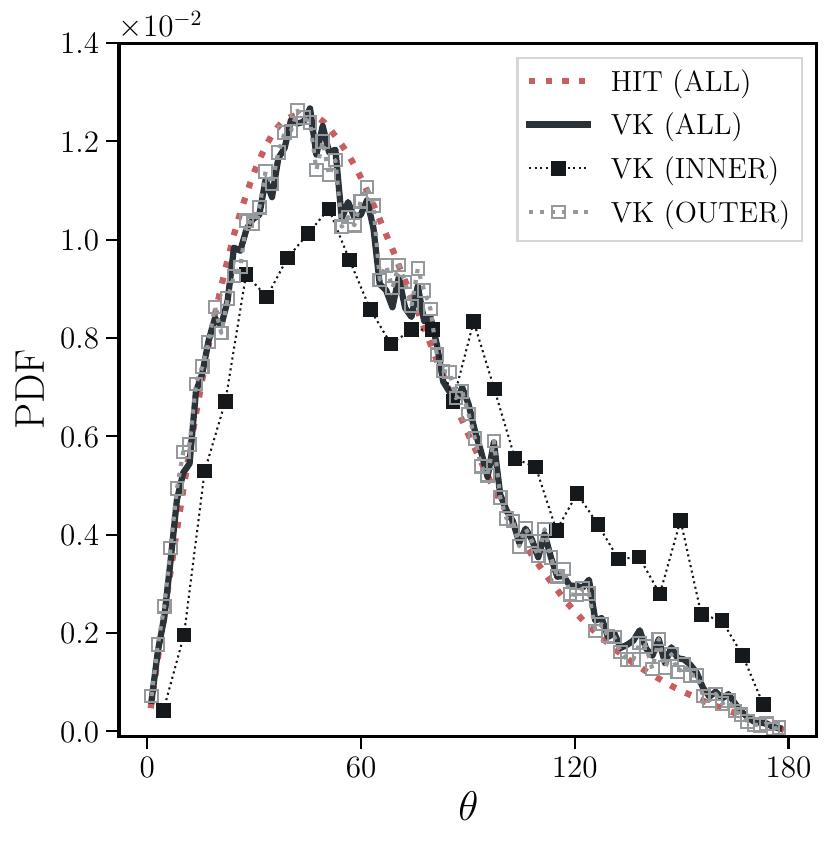}
    \caption{Probability density functions of the pair dispersion angle during the superdiffusive regime, for particles in the entire domain and in the inner and outer subregions indicated in Fig.~\ref{fig:subregs}. From left to right, HIT, TG, and VK datasets. The PDF of the HIT dataset in the entire domain is shown in the three panels as a reference.}
    \label{pda_histogram_innout}
\end{figure}

Finally, we consider how particle pairs behave in the inner and outer subregions. The first includes the central saddle point of the VK and TG cells, and the second does not, while at the same time both are large enough to allow for analysis of the experimental data. Note that in the VK experiment the number of trajectory pairs after conditioning to subregions becomes significantly smaller. Thus, we compare only the APDA averaged over all initial particle separations in the superdiffusive regime (Table \ref{tab:averages}). We show in Fig.~\ref{pda_histogram_innout} the PDFs of $\theta$ within the superdiffusive regime, in the entire cells as well as in the inner and outer subregions, both for the TG and VK flows. As HIT is statistically homogeneous, we also show the same analysis for this flow as differences between the subregions provide an estimation of uncertainties in $\theta$ and in the APDA.

\begin{table}
\caption{Average pair dispersion angle computed in the plateau (i.e., in the superdiffusive regime) for the entire cell, as well as for the inner and outer subregions indicated in Fig.~\ref{fig:subregs} for HIT, TG, and VK flows. As the HIT flow is expected to be homogeneous, differences between the subregions provide an estimation of uncertainties in the APDA.}
\label{tab:averages}
\begin{ruledtabular}
    \begin{tabular}{lcccccc}
             && \textbf{ALL} && \textbf{INNER} && \textbf{OUTER} \\\hline
        \textbf{HIT}  && {$59.6\degree$} && {$61.6\degree$} && {$59.5\degree$} \\
        \textbf{TG}   && {$60.4\degree$} && {$64.4\degree$} && {$59.5\degree$} \\
        \textbf{VK}   && {$62.2\degree$} && {$75.0\degree$} && {$61.5\degree$}
    \end{tabular}
\end{ruledtabular}
\end{table}

The PDFs of the HIT dataset in Fig.~\ref{pda_histogram_innout} show no significant differences in $\theta$ between the entire domain and the two subregions, and the APDA in table \ref{tab:averages} varies within $\approx 1^\circ$. A very different behavior is seen in the TG and VK datasets. In both flows the central subregion has a larger APDA than in HIT (see table \ref{tab:averages}) and the PDFs in that subregion display a shift towards larger values of $\theta$ (see Fig.~\ref{pda_histogram_innout}), while the PDFs in the entire cell or in the outer region are closer to the PDFs of HIT as shown before in Fig.~\ref{pda_histogram}. As also seen in the center and corner subregions, the presence of the saddle point seems to strongly affect the pair dispersion angle, increasing the mean angle between particles.

\subsection{Low-order models for the Taylor-Green and von K\'arm\'an mean flows}\label{subsec:low_order_models}

In order to better understand the effect of the saddle point on the dynamics of the pair dispersion process, we now study how the APDA of particle pairs embedded on a saddle node behaves. We first consider a simple saddle node, and later we consider more detailed analytical expressions for the specific saddle nodes located at the center of the mean TG and VK flows. The simplest stationary saddle flow can be described by $\dot{x} = \lambda x$, $\dot{y} = \lambda y$ and $\dot{z} = -2\lambda z$, which verifies mass conservation as the sum of the eigenvalues is equal to zero, $\lambda + \lambda - 2\lambda = 0$. The trajectory of a particle in this flow is given by $x(t) = x_0 \, \rm{exp}(\lambda t)$, $y(t) = y_0 \, \rm{exp}(\lambda t)$, and $z(t) = z_0 \, \rm{exp}(-2\lambda t)$. Using Eq.~(\ref{eq:angle}), and picking two particles with initial positions $(x_1, y_1, z_1)$ and $(x_2, y_2, z_2)$, the pair dispersion angle in this saddle flow is then given by
\begin{equation}
    \rm{cos}(\theta) = \frac{\rm{sign}(\lambda) (1 - 2 \alpha^2 e^{-6\lambda t})}{(1 + c^2 e^{-6\lambda t})^{1/2}(1 + 4 \alpha^2 e^{-6\lambda t})^{1/2}},
    \label{eq:saddle}
\end{equation}
where $\alpha^2 = (z_2-z_1)^2[(x_2-x_1)^2+(y_2-y_1)^2]^{-1}$. For times satisfying $t \gg \lambda^{-1}$, the right-hand side approaches 1, leading to a pair dispersion angle that tends toward zero. This confirms that the saddle-point flow enforces alignment in the pair dispersion process. We can therefore expect that particle pairs in the vicinity of a saddle point will be characterized by a smaller APDA.

\begin{figure}
    \centering
    \includegraphics[width=.9\linewidth]{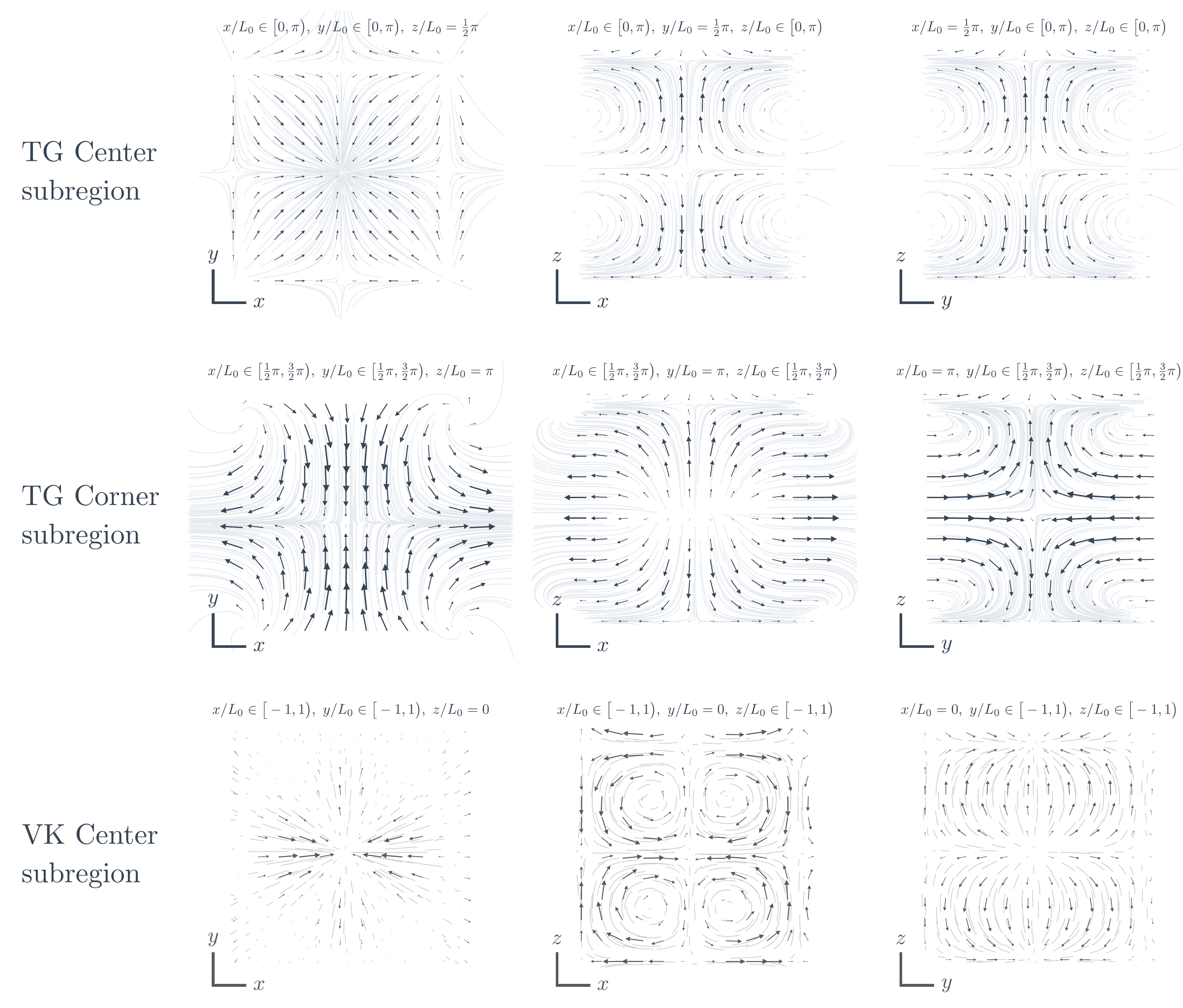}
    \caption{Mean TG and VK flows in different two-dimensional slices of the center and corner subregions, reconstructed using a few modes of analytical models of the TG \cite{TaylorGreen1937} and VK flows \cite{Berning2023}.
		}
    \label{fig:topology}
\end{figure}

We next perform a more detailed analysis of the effect on saddle points on the APDA using analytical expressions which are specific for the mean large-scale TG and VK flows. To model the large-scale flow in the TG case, we take the first two modes of the expansion by Taylor and Green \cite{TaylorGreen1937}, which are enough to capture the main and secondary circulation. These are given by a superposition of sines and cosines with two wave numbers ($k_0$ and $2k_0$). The resulting flow, computed in the center and corner subregions, is illustrated in Fig.~\ref{fig:topology}. For the VK flow we use the three-modes model in \cite{Berning2023}, based on a reconstruction of the flow from experimental observations in a cylindrical vessel (even though our experiment has a square section, we focus here on the center subregion with the saddle point, and assume that the central saddle point has similar properties independently of the section of the domain). Two modes in \cite{Berning2023} correspond to the secondary poloidal circulation, while one mode accounts for the primary circulation and consists of vortices in the $xy$ plane with shear in the $z$ direction. Figure \ref{fig:topology} shows also a reconstruction using these modes for the center subregion of the VK flow in three different slices.

\begin{figure}
    \includegraphics[width=.6\textwidth]{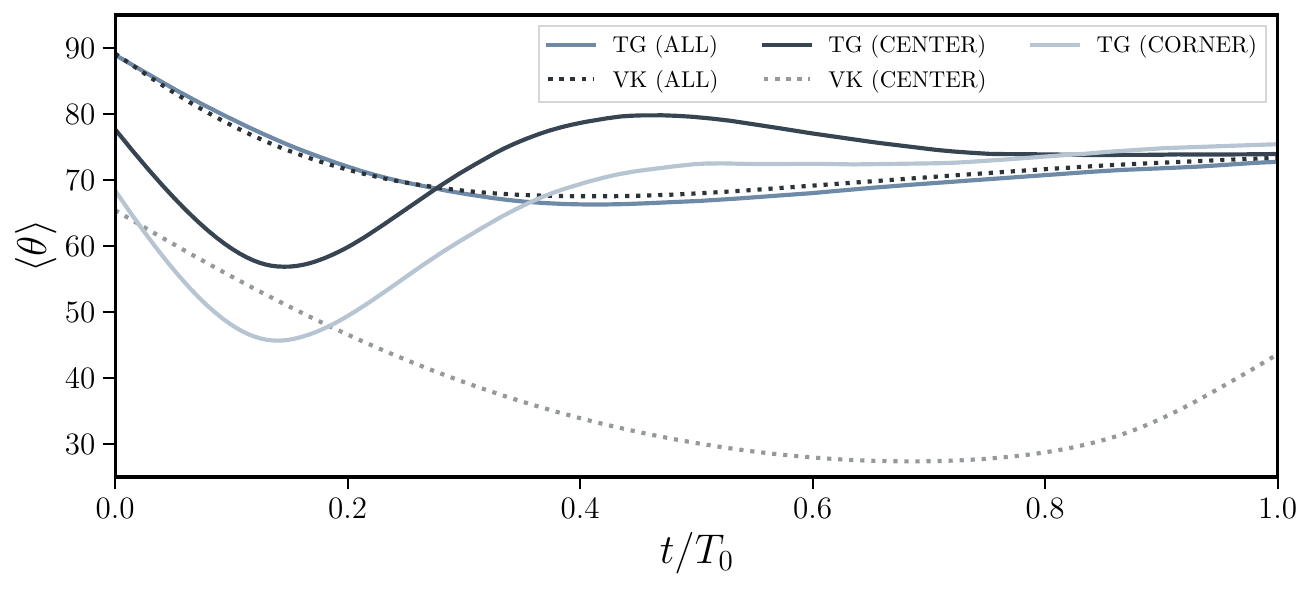}
    \caption{APDA for particles in the low-order analytical models of the TG and VK flows. Integration of pairs of particles is shown for the entire flow (``ALL''), and for the center and corner subregions (in the later case, only for the TG flow). The initial separation for all computed pairs of particles was set to $\Delta r_0/L = 10^{-4}$.}
    \label{fig:pda_models}
\end{figure}

For each of these analytical models, we performed a fourth-order Runge-Kutta integration of pairs of particles, starting in random positions in the displayed subregions in Fig.~\ref{fig:topology}, with initial distances ranging from $\Delta r_0/L_0 = 10^{-4}$ to $10^{-1}$. The results of the evolution of the APDA for these subregions, and for pairs of particles integrated in the entire low-order reconstructed flows, are shown in Fig.~\ref{fig:pda_models}. As no significant differences were observed within the considered range of initial particles' separations, we show results only for $\Delta r_0/L = 10^{-4}$. 

Pairs of particles starting in the center TG and VK subregions, and in the TG corner subregions, reach a minimum in the APDA. For the corner TG and center VK subregions (which have a saddle point with similar properties), a minimum in the APDA between $\approx 45^\circ$ and $30^\circ$ degrees is reached, while for the center TG subregion the minimum is at $\approx 60^\circ$. These values are comparable with the values of the minima reported for the turbulent flows in Sec.~\ref{sec:local}. The times to reach these minima are $t/T_0 \approx 0.2$ in the TG flow, and much larger ($\approx 0.7$) in the reconstructed VK flow. In spite of this, already at $t/T_0 \approx 0.2$ the saddle in the center of the reconstructed VK flow generates values of the APDA close to $45^\circ$. For long times the APDA in most of these low-order reconstructed flows tend to $\approx 70^\circ$, also in agreement with some of the observations in Fig.~\ref{fig:tg_center_corner}. In other words, the rate at which the alignment in the pair dispersion process is lost seems to be not only affected by the turbulence, but also locally increased by the effect that some of these saddle points have on the particles.

These results are interesting and somewhat unexpected in the light of several studies indicating that the VK flow, under both Eulerian and Lagrangian metrics, behaves similarly to HIT when only the central region of the domain close to the shear layer is considered \cite{Mordant_2004, Voth1998, Mordant1997}. The study of the APDA seems to indicate instead that the persistence of a saddle point in that region has a strong impact in the turbulent transport of tracers, and probably on other metrics.

\section{\label{sec:level4}Conclusions}

We examined the behavior of the dispersion of particle pairs using the averaged pair dispersion angle introduced in \cite{Shnapp2023}, in turbulent flows with varying degrees of anisotropy and inhomogeneity. To that end we used data from an experimental von K\'arm\'an flow, and from numerical simulations of a Taylor-Green flow as well as of homogeneous and isotropic turbulence. The metric used in the analysis quantifies the mean level of alignment between the particles' separation vector and their relative velocity. The ballistic and superdiffusive regimes of pair dispersion can be identified in the three datasets using this metric. They correspond respectively to a rapid decrease of the pair dispersion angle from an initial value of $90 \degree$, followed by a stabilization of this angle at $\approx 60 \degree$. The later value was claimed to be universal for the superdiffusive turbulent regime \cite{Shnapp2023}. When considering trajectories in the entire simulation or experimental domain, we obtain values of the angle compatible with this value for the three datasets, irrespective of the flow considered. This indicates that the pair dispersion angle can be used as a metric even in the presence of flow anisotropies, highlighting its utility as a tool for studying turbulent transport in complex, real-world flows.

However, the von K\'arm\'an and Taylor-Green flows also display distinct features that separates them from homogeneous and isotropic turbulence. First, the pair dispersion angle in the ballistic regime reaches a pronounced minimum in the Von K\'arm\'an experiment. Second, when studying pair dispersion in subregions of these flows the behavior of the pair dispersion angle is affected both in the ballistic and in the superdiffusive regimes. Our study reveals that the presence of long-lived saddle points in these flows significantly influences pair dispersion, leading to deviations from the behavior observed in homogeneous turbulence. In particular, saddle points and local geometrical properties of the mean flow cause stronger alignment of particle pairs during the ballistic regime, and faster decorrelation in the diffusive regime. These effects are most pronounced in the central region of the von K\'arm\'an and Taylor-Green flows, where the influence of mean large-scale structures is strongest.

Our findings indicate that local flow geometry plays a critical role in determining particle transport behavior, and should be considered in models of turbulent mixing in both natural and industrial applications. This result strengthens previous studies indicating that particle behavior is strongly affected by null points of the velocity field \cite{Mora_2021}, by flow non-stationarity \cite{Zapata_2024}, and by the flow topology \cite{Yokoi_1993, Angriman2020}. Finally, the results indicate that even though the central region of the von K\'arm\'an experiment is often used to study homogeneous and isotropic turbulence \cite{Mordant_2004, Voth1998, Mordant1997}, as a result of its strong shear and local high Reynolds number, some of its transport properties may differ significantly from those found in homogeneous flows.

\begin{acknowledgments}
The authors acknowledge financial support from UBACyT Grant No.~20020220300122BA, and from Proyecto REMATE of the Redes Federales de Alto Impacto, Argentina.
B.L.E and M.N. contributed equally to this work.
\end{acknowledgments}

\bibliography{main}

\end{document}